\documentclass[%
 aip,
 sd,%
 amsmath,amssymb,
preprint,%
]{revtex4-1}

\usepackage{graphicx}
\usepackage{dcolumn}
\usepackage{bm}

\usepackage{graphicx,epic,amscd,amssymb,amsmath,multirow}
\usepackage[all]{xy}

\newcommand{\DIS}{\displaystyle}
\newcommand{\qed}{\hfill $\Box$}
\def\C{{\mathbb C}}
\def\Z{{\mathbb Z}}
\def\P{{\mathbb P}}
\def\Q{{\mathbb Q}}
\def\T{{\mathbb T}}
\def\A{{\mathbb A}}
\def\R{{\mathbb R}}
\def\nn{{\nonumber}}

\def\bx{\mbox{\boldmath $x$}}
\def\by{\mbox{\boldmath $y$}}
\def\ch{{\rm ch}\mkern2mu}
\def\sh{{\rm sh}\mkern2mu}

\newtheorem{thm}{Theorem}
\newtheorem{corollary}{Corollary}
\newtheorem{remark}{Remark}
\newtheorem{example}{Example}

\newtheorem{proposition}{Proposition}

\begin{document}

\preprint{
}

\title[Generators of rank 2 cluster algebras of affine types]{Generators of rank 2 cluster algebras of affine types via linearization of seed mutations}

\author{Atsushi Nobe}
 \email{nobe@faculty.chiba-u.jp}
\affiliation{ 
Faculty of Education, Chiba University,\\ 1-33 Yayoi-cho Inage-ku, Chiba 263-8522, Japan
}%

\date{\today}

\begin{abstract}
From the viewpoint of integrable systems on algebraic curves, we discuss linearization of birational maps arising from the seed mutations of types $A^{(1)}_1$ and $A^{(2)}_2$, which enables us to construct the set of all cluster variables generating the corresponding cluster algebras.
These birational maps respectively induce discrete integrable systems on algebraic curves referred to as the types of the seed mutations from which they are arising.
The invariant curve of type $A^{(1)}_1$ is a conic, while the one of type $A^{(2)}_2$ is a singular quartic curve.
By applying the blowing-up of the singular quartic curve, the discrete integrable system of type $A^{(2)}_2$ on the singular curve is transformed into the one on the conic, the invariant curve of type $A^{(1)}_1$.
We show that the both discrete integrable systems of types $A^{(1)}_1$ and $A^{(2)}_2$ commute with each other on the conic, the common invariant curve.
We moreover show that these integrable systems are simultaneously linearized by means of the conserved quantities and their general solutions are respectively obtained.
By using the general solutions, we construct the sets of all cluster variables generating the cluster algebras of types $A^{(1)}_1$ and $A^{(2)}_2$, respectively.
%
\end{abstract}

\maketitle

\section{Introduction}
\label{sec:intro}

Seed mutations in cluster algebras, which produce new cluster variables from old ones in terms of their birational equations called the exchange relations, can be regarded as time evolutions of dynamical systems governed by birational maps.
Appropriate choices of the directions of seed mutations in adequate cluster algebras lead to proper dynamical systems; discrete integrable systems.
In fact, since the introduction of cluster algebras by Fomin and Zelevinsky in 2002 \cite{FZ02} we have found many applications of cluster algebras in the field of discrete or quantum integrable systems such as discrete soliton equations, integrable maps on algebraic curves, discrete/$q$- Painlev\'e equations and $Y$-systems \cite{FZ03-2,IIKNS10,IIKKN13,IIKKN13-2,Okubo13,Mase13,Okubo15,Marshakov13,Mase16,Nobe16,BGM18}.
The number of cluster variables in the initial seed is called the rank of the cluster algebra.
There are three types of cluster algebras of rank 2; finite, affine and strictly hyperbolic types \cite{Kac94}. 
The type of a cluster algebra is referred to as the type of the generalized Cartan matrix called the Cartan counterpart of the exchange matrix \cite{FZ02}.
Seed mutations in rank 2 cluster algebras of finite and affine types lead to two-dimensional discrete integrable systems \cite{Nobe16}, as will be shown later.
Since a cluster algebra of finite type has finite periodicity, its cluster variables are easily obtained.
On the other hand, to construct the set of all cluster variables generating a cluster algebra of non-finite type is not so easy since the algebra has infinite periodicity.
In fact, we find that infinitely many rank 2 cluster algebras of strictly hyperbolic type lead to chaotic dynamical systems, hence, it seems difficult to obtain all cluster variables of such cluster algebras. (We will report on this subject in a forthcoming paper.)
Nevertheless, we can explicitly construct the set of all cluster variables generating rank 2 cluster algebras of affine types in terms of their integrable structures. 
Although rank two cluster algebras of affine types and their generators have already been investigated intensively \cite{SZ04, CZ06, Zelevinsky07}, their methods based on combinatorics are different from ours based on geometry of seed mutations.

It is well known that, among discrete integrable systems, a paradigmatic family of  two-dimensional birational maps called the QRT maps plays an important role \cite{QRT89,Tsuda04}.
Actually, many reductions of discrete soliton equations and many autonomous limits of discrete Painlev\'e equations are members of the family of QRT maps. 
Moreover, since QRT maps are generated by point additions on elliptic surfaces, connection with a QRT map gives a geometric interpretation of the discrete system.
Therefore, it is necessary to investigate cluster algebras of rank 2, some of which are directly connected with QRT maps, thoroughly in order to grasp integrable structures of cluster algebras of higher rank.
In the preceding paper \cite{Nobe16}, we investigated cluster algebras of rank 2 from the viewpoint of discrete integrable systems on plane curves, and a direct connection between the seed mutations and the discrete Toda lattice both of which are of type $A^{(1)}_1$ was established.
Namely, we showed that the seed mutations of type $A^{(1)}_1$ are naturally regarded as degenerate limits of additive group actions on an elliptic curve equivalent to the time evolutions of the discrete Toda lattice of the same type.
In this paper, we further investigate rank 2 cluster algebras of affine types, namely, of types $A^{(1)}_1$ and $A^{(2)}_2$, and show linearization of their seed mutations.
We first reduce birational maps from seed mutations in these cluster algebras.
We also refer to the birational maps and to the dynamical systems governed by them as the types of the seed mutations from which they are arising.
The invariant curve of type $A^{(1)}_1$ is a conic, while the one of type $A^{(2)}_2$ is a singular quartic curve.
Notwithstanding, we can transform the singular curve of type $A^{(2)}_2$ into the non-singular conic of type $A^{(1)}_1$ by means of the blowing-up of the singular curve.
Then the birational map of type $A^{(2)}_2$ on the singular curve is simultaneously transformed into the one on the conic.
Consequently, we find that these birational maps on the conic commute with each other and are simultaneously linearized in terms of the conserved quantities.
We then solve the integrable systems governed by the birational maps of both types, and obtain the general solutions to their initial value problems, respectively.
It is straightforward to construct the sets of all cluster variables generating the cluster algebras of types $A^{(1)}_1$ and $A^{(2)}_2$ from the general solutions.

This paper is organized as follows.
In \S \ref{sec:CAQRT}, we briefly review cluster algebras.
We then introduce seed mutations of type $A^{(2)}_2$ and reduce a dynamical system governed by a quartic birational map from the seed mutations of type $A^{(2)}_2$.
We find that the invariant curve of the dynamical system is a singular quartic curve.
It follows that the dynamical system of type $A^{(2)}_2$ is integrable in the sense of Liouville.
We moreover resolve the singularity by blowing-up the curve, and get a conic as the strict transform of the singular curve.
Simultaneously, we obtain a discrete integrable system on the conic governed by a cubic birational map, which is conjugate to the quartic map of type $A^{(2)}_2$ with respect to the blowing-up.
In \S \ref{sec:A11A22}, we first show that the conic thus obtained is nothing but the invariant curve of the QRT map arising from the seed mutations of type $A^{(1)}_1$.
We then show the commutativity of the birational maps of types $A^{(1)}_1$ and $A^{(2)}_2$ on the conic.
These birational maps are simultaneously linearized by means of the conserved quantities, and the initial value problems of the dynamical systems governed by these maps are respectively solved.
We finally construct the sets of all cluster variables generating the cluster algebras of types $A^{(1)}_1$ and $A^{(2)}_2$ in terms of these general solutions, respectively.
\S \ref{sec:CONCL} is devoted to concluding remarks.
In Appendix \ref{sec:App}, we give a method to find the invariant curve of type $A^{(2)}_2$ with the aid of tropical geometry.

\section{Seed mutations of type $\boldsymbol{A^{(2)}_2}$ and birational maps}
\label{sec:CAQRT}
\subsection{Cluster algebras}
\label{subsec:CA}

We briefly review a portion of cluster algebras \cite{FZ02,FZ03,FZ07}.
Let $\bx=(x_1,x_2,\ldots,x_n)$ be the set of generators of the ambient field $\mathcal{F}=\mathbb{QP}(\bx)$, where $\P=\left(\P,\cdot,\oplus\right)$ is a semifield endowed with multiplication $\cdot$ and auxiliary addition $\oplus$ and $\mathbb{QP}$ is the group ring of $\P$ over $\Q$.
Also let $\by=(y_1,y_2,\ldots,y_n)$ be an $n$-tuple in $\P^n$ and $B=(b_{ij})$ be an $n\times n$ skew-symmetrizable integral matrix.
The triple $(\bx,\by,B)$ is referred to as the seed.
We also refer to $\bx$ as the cluster of the seed, to $\by$ as the coefficient tuple and to $B$ as the exchange matrix.
Elements of $\bx$ and $\by$ are called cluster variables and coefficients, respectively.

We introduce seed mutations.
Let $k\in[1,n]$ be an integer.
The seed mutation $\mu_k$ in the direction $k$ transforms a seed $(\bx,\by,B)$ into the seed $(\bx^\prime,\by^\prime,B^\prime):=\mu_k(\bx,\by,B)$ defined by the following birational equations in the seeds called the exchange relations:
\begin{align}
b_{ij}^\prime
&=
\begin{cases}
-b_{ij}&\mbox{$i=k$ or $j=k$},\\
b_{ij}+[-b_{ik}]_+b_{kj}+b_{ik}[b_{kj}]_+&\mbox{otherwise},\\
\end{cases}
\label{eq:mutem}\\
y_j^\prime
&=
\begin{cases}
y_k^{-1}&\mbox{$j=k$},\\
y_jy_k^{[b_{kj}]_+}(y_k\oplus 1)^{-b_{kj}}&\mbox{$j\neq k$},\\
\end{cases}
\label{eq:mutcoef}\\
x_j^\prime
&=
\begin{cases}
\DIS\frac{y_k\prod x_i^{[b_{ik}]_+}+\prod x_i^{[-b_{ik}]_+}}{(y_k\oplus 1)x_k}&\mbox{$j=k$},\\
x_j&\mbox{$j\neq k$},\\
\end{cases}
\label{eq:mutcv}
\end{align}
where we define $[a]_+:=\max[a,0]$ for $a\in\Z$.

Let $\T_n$ be the $n$-regular tree whose edges are labeled by $1, 2, \ldots, n$ so that the $n$ edges emanating from each vertex receive different labels.
We write $t\ \overset{k}{\begin{xy}\ar @{-}(10,0)\end{xy}}\ t^\prime$ to indicate that vertices $t,t^\prime\in\T_n$ are joined by an edge labeled by $k$.
We assign a seed $\Sigma_t=(\bx_t,\by_t,B_t)$ to every vertex $t\in\T_n$ so that the seeds assigned to the endpoints of any edge $t\ \overset{k}{\begin{xy}\ar @{-}(10,0)\end{xy}}\ t^\prime$ are obtained from each other by the seed mutation in direction $k$.
We refer to the assignment $\T_n\ni t\mapsto\Sigma_t$ as a cluster pattern.
We write the elements of $\Sigma_t$ as follows
\begin{align*}
\bx_t=(x_{1;t},\ldots,x_{n;t}),\quad
\by_t=(y_{1;t},\ldots,y_{n;t}),\quad
B_t=(b_{ij}^t).
\end{align*}

Given a cluster pattern $\T_n\ni t\mapsto\Sigma_t$, we denote the union of clusters of all seeds in the pattern by
\begin{align*}
\mathcal{X}
=
\bigcup_{t\in\T_n}\bx_t
=
\left\{
x_{i;t}\ |\ t\in\T_n,\ 1\leq i\leq n
\right\}.
\end{align*}
The cluster algebra $\mathcal{A}$ associated with a given cluster pattern is the $\mathbb{ZP}$-subalgebra of the ambient field $\mathcal{F}$ generated by all cluster variables: $\mathcal{A}=\mathbb{ZP}[\mathcal{X}]$.
We refer to the set $\mathcal{X}$ of all cluster variables as the set of generators of $\mathcal{A}$.
Remark that $\mathcal{A}$ is also generated by its initial cluster variables $\bx$ as a Laurent polynomial subring of the ambient field $\mathcal{F}$ \cite{FZ02}.

\subsection{Seed mutations of type $\boldsymbol{A^{(2)}_2}$}
\begin{figure}[htbp]
\centering
$
\xymatrix{{t_{-2}}\ar @{-}(3,-2);(13,-12)^1&&t_0&\ar @{-}(32,-2);(42,-12)^1&t_2&\ar @{--}(60,-2);(70,-12)^1\\
\ar @{--}(-13,-12);(-3,-2)^2&{t_{-1}}\ar @{-}(19,-12);(29,-2)^2&\ar @{-}(46,-12);(56,-2)^2&t_1\\}
$
\caption{
The regular binary tree $\T_2$ to which the cluster pattern $\T_2\ni t_m\mapsto\Sigma_m=\left(\bx_m,\by_m,B_m\right)$ is assigned.
}
\label{fig:binarytree}
\end{figure}
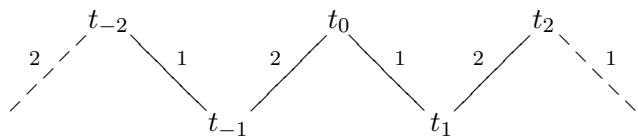

Let us consider the cluster algebra $\mathcal{A}$ generated from the following initial seed $\Sigma_0=\left(\bx_0,\by_0,B_0\right)$
\begin{align}
\bx_0=\left(x_1,x_2\right),\quad
\by_0=\left(y_1,y_2\right),\quad
B_0=\left(\begin{matrix}0&-4\\1&0\\\end{matrix}\right),
\label{eq:initialseed}
\end{align}
where $\bx_0$ is the cluster, $\by_0$ is the coefficient tuple and $B_0$ is the exchange matrix. 
The semifield $\P$ is arbitrarily chosen.
Let  $\T_2$  be the regular binary tree whose edges are labeled by the numbers 1 and 2. 
Noting that a seed mutation is an involution (see the exchange relations \eqref{eq:mutem}-\eqref{eq:mutcv}), we obtain the tree $\T_2$ as an infinite chain.
We consider the cluster pattern $\T_2\ni t_m\mapsto\Sigma_m=\left(\bx_m,\by_m,B_m\right)$ in figure \ref{fig:binarytree}, and fix it throughout this paper.

We denote the clusters, the coefficients and the exchange matrices by
\begin{align}
\bx_{2k}=(x_{1;2k},x_{2;2k})
&\overset{\mu_1}{\longleftrightarrow}&
\bx_{2k+1}=(x_{1;2k+1},x_{2;2k+1})
\overset{\mu_2}{\longleftrightarrow}&&
\bx_{2k+2}=(x_{1;2k+2},x_{2;2k+2}),
\label{eq:mu1mu2x}\\
\by_{2k}=(y_{1;2k},y_{2;2k})
&\overset{\mu_1}{\longleftrightarrow}&
\by_{2k+1}=(y_{1;2k+1},y_{2;2k+1})
\overset{\mu_2}{\longleftrightarrow}&&
\by_{2k+2}=(y_{1;2k+2},y_{2;2k+2}),
\label{eq:mu1mu2y}\\
B_{2k}=\left(b_{ij}^{2k}\right)
&\overset{\mu_1}{\longleftrightarrow}&
B_{2k+1}=\left(b_{ij}^{2k+1}\right)
\overset{\mu_2}{\longleftrightarrow}&&
B_{2k+2}=\left(b_{ij}^{2k+2}\right)
\label{eq:mu1mu2B}
\end{align}
for $k\in\Z$, respectively.

We see from the exchange relation \eqref{eq:mutem} that the exchange matrices have period two:
\begin{align*}
B_m
=
\begin{cases}
B_0&\mbox{$m$ even,}\\
-B_0&\mbox{$m$ odd.}\\
\end{cases}
\end{align*}
It follows that we have the following Cartan counterpart $A(B_m)$ of $B_m$ \cite{FZ03}:
\begin{align*}
A(B_m)
:=
\left(2\delta_{ij}-\left|b_{ij}^m\right|\right)
=
\left(\begin{matrix}2&-4\\-1&2\\\end{matrix}\right)
\quad
\mbox{for ${}^\forall m\in\Z$}.
\end{align*}
Since $A(B_m)$ is the generalized Cartan matrix of type $A^{(2)}_2$, we refer to the cluster algebra $\mathcal{A}$ as of type $A^{(2)}_2$.
The seed mutations $\mu_1$ and $\mu_2$ are also referred to as of type $A^{(2)}_2$.
Note that there is no quiver representation of $B_m$ since it is not skew-symmetric but skew-symmetrizable \cite{FZ02}.

The cluster pattern $\T_2\ni t_m\mapsto\Sigma_m=\left(\bx_m,\by_m,B_m\right)$ of the cluster algebra $\mathcal{A}$ of type $A^{(2)}_2$ induces a dynamical system governed by a quartic birational map on the projective plane $\P^2(\C)$.
We often refer to the birational map and to the dynamical system as of type $A^{(2)}_2$ as well as the seed mutations from which they are arising.
\begin{proposition}
Let $\mathcal{A}$ be the cluster algebra of type $A^{(2)}_2$.
For $t\geq0$, we associate $z^t$ and $w^t$ with the seed of $\mathcal{A}$ as
\begin{align*}
z^t
=
\frac{x_{1;2t}}{\sqrt[4]{y_{2;2t}}},
\quad
w^t
=
y_{1;2t}x_{2;2t}.
\end{align*}
Assume $z^0,w^0\in\P^2(\C)$.
Then the cluster pattern $\T_2\ni t_m\mapsto\Sigma_m=\left(\bx_m,\by_m,B_m\right)$ of $\mathcal{A}$ induces a birational map $\psi_{\rm vh}:$
\begin{align}
(z^t,w^t)\mapsto(z^{t+1},w^{t+1})
=
\left(
\frac{w^t+1}{z^t},
\frac{\left(z^{t+1}\right)^4+1}{w^t}
\right)
\label{eq:bmA22}
\end{align}
on $\P^2(\C)$ from the initial seed $\Sigma_0=\left(\bx_0,\by_0,B\right)$.
\end{proposition}

(Proof)\quad
From the exchange relation \eqref{eq:mutcoef}, it immediately follows the equalities among the coefficients:
\begin{align*}
&y_{1;2k}y_{1;2k+1}=1,
\\
&y_{2;2k+1}
=
y_{2;2k}y_{1;2k}^{[-4]_+}(y_{1;2k}\oplus 1)^4
=
y_{2;2k}(y_{1;2k}\oplus 1)^4,
\\
&y_{2;2k+1}y_{2;2k+2}=1,
\\
&y_{1;2k+2}
=
y_{1;2k+1}y_{2;2k+1}^{[-1]_+}(y_{2;2k+1}\oplus 1)^1
=
y_{1;2k+1}(y_{2;2k+1}\oplus 1).
\end{align*}
Moreover, by using the exchange relation \eqref{eq:mutcv} for the cluster variables and the above equalities in the coefficients, we compute
\begin{align}
x_{1;2k+1}
&=
\frac{y_{1;2k}x_{2;2k}+1}{(y_{1;2k}\oplus 1)x_{1;2k}}
=
\frac{y_{1;2k}x_{2;2k}+1}{\sqrt[4]{\frac{1}{y_{2;2k}y_{2;2k+2}}}x_{1;2k}},
\nn\\
x_{2;2k+1}
&=x_{2;2k},
\nn\\
x_{1;2k+2}
&=
x_{1;2k+1}
=
\frac{y_{1;2k}x_{2;2k}+1}{\sqrt[4]{\frac{1}{y_{2;2k}y_{2;2k+2}}}x_{1;2k}},
\label{eq:x1te}\\
x_{2;2k+2}
&=
\frac{y_{2;2k+1}(x_{1;2k+1})^4+1}{(y_{2;2k+1}\oplus 1)x_{2;2k+1}}
=
\frac{\frac{1}{y_{2;2k+2}}(x_{1;2k+2})^4+1}{y_{1;2k}y_{1;2k+2}x_{2;2k}}.
\label{eq:x2te}
\end{align}

By setting
\begin{align}
z^t
=
\frac{x_{1;2t}}{\sqrt[4]{y_{2;2t}}},
\quad
w^t
=
y_{1;2t}x_{2;2t},
\label{eq:zwxy}
\end{align}
we obtain the birational map
\begin{align*}
\psi_{\rm vh}:
(z^t,w^t)\mapsto(z^{t+1},w^{t+1})
=
\left(
\frac{w^t+1}{z^t},
\frac{(z^{t+1})^4+1}{w^t}
\right)
\end{align*}
from the exchange relations \eqref{eq:x1te} and \eqref{eq:x2te}.
Note that the map $\psi_{\rm vh}$ is equivalent to the application of successive seed mutations $\mu_1$ and $\mu_2$.
\qed

We find that the birational map $\psi_{\rm vh}$ of type $A^{(2)}_2$ is integrable in the sense of Liouville.
Actually, the two-dimensional map $\psi_{\rm vh}$ has an invariant curve parametrized with a conserved quantity depending on the initial point $(z^0,w^0)$.
The invariant curve is concretely constructed as follows.

\subsection{Invariant curves}
Let $\overline\gamma_\lambda$ be the curve on the projective plane $\P^2(\C)$ defined to be
\begin{align}
\begin{cases}
\overline\gamma_\lambda
:=
\gamma_\lambda\cup\left\{P_\infty^\prime\right\},
\\
\gamma_\lambda
:=
\left(f(z,w)=0\right),
\\
f(z,w):=(w+1)^2+\lambda z^2w+z^4,
\end{cases}
\label{eq:ICA22}
\end{align}
where the point $P_\infty^\prime$ at infinity is given by $[0:1:0]$ in the homogeneous coordinate $(z,w)\mapsto[z:w:1]$ and the curve is parametrized with $\lambda\in\P^1(\C)$.
The curve $\gamma_\lambda$ is the affine part of $\overline\gamma_\lambda$.
The base points of the pencil $\left\{\overline\gamma_\lambda\right\}_{\lambda\in\P^1(\C)}$ are the following 6 points:
\begin{align*}
P_\infty^\prime=[0:1:0],
\quad
\mathfrak{p}:=[0:-1:1],
\quad
\left[\zeta_8^{i}:0:1\right]
\quad
(i=1,3,5,7),
\end{align*}
where $\zeta_8$ is the eighth root of 1.
Let $\overline{\mathcal{B}}$ and $\mathcal{B}$ be the sets of the base points:
\begin{align}
\begin{cases}
\overline{\mathcal{B}}
:=
\mathcal{B}
\cup
\left\{P_\infty^\prime\right\},
\\
\mathcal{B}
:=
\left\{
\mathfrak{p}
\right\}
\cup\left\{
\left.
\left[\zeta_8^{i}:0:1\right]\ \right|\
i=1,3,5,7
\right\}.
\end{cases}
\label{eq:setofbasepoints}
\end{align}

\begin{figure}[b]
\centering
\includegraphics[scale=.75]{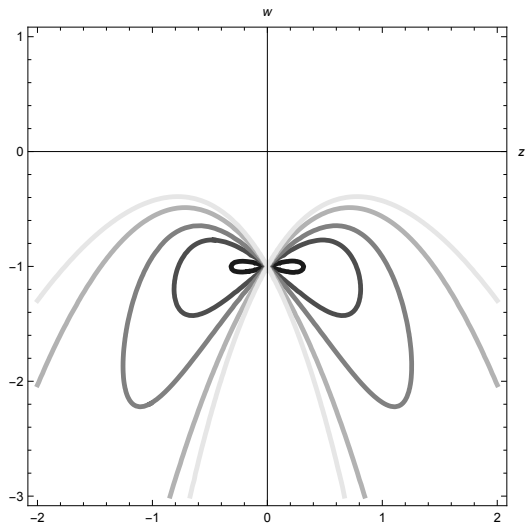}
\caption{Members of the pencil $\left\{\overline\gamma_\lambda\right\}_{\lambda\in\P^1(\C)}$ for $\lambda=0.1$, $0.6$, $1.1$, $2.1$ and $3.1$.
A curve with smaller $\lambda$ is colored darker.
Each curve passes through the base point $\mathfrak{p}=(0,-1)$.}
\label{fig:pencil}
\end{figure}

\begin{remark}
The curve $\overline\gamma_\lambda$ is a singular quartic curve which has the singularity at the point $\mathfrak{p}=(0,-1)$.
The singular point $\mathfrak{p}$ is an ordinary double point and is the base point of the pencil $\left\{\overline\gamma_\lambda\right\}_{\lambda\in\P^1(\C)}$ as well (see figure \ref{fig:pencil}).
\end{remark}

The singular curve $\overline\gamma_\lambda$ is nothing but the invariant curve of the dynamical system governed by the birational map $\psi_{\rm vh}$ of type $A^{(2)}_2$.
\begin{thm}\label{thm:A22ds}
Assume $(z^0,w^0)$ to be a point on $\P^2(\C)-\overline{\mathcal{B}}$.
Let $S:=\left\{(z^t,w^t)\right\}_{t\geq0}$ be a sequence of points on $\P^2(\C)$ generated from $(z^0,w^0)$ by applying  the birational map $\psi_{\rm vh}$ of type $A^{(2)}_2$, repeatedly.
Then any point in $S$ is on the affine curve $\gamma_\lambda$ for
\begin{align}
\lambda
=
-\frac{(w^0+1)^2+(z^0)^4}{(z^0)^2w^0}.
\label{eq:lambda}
\end{align}
\end{thm}

(Proof)\quad
Since $(z^0,w^0)$ is not a base point of the pencil $\left\{\overline\gamma_\lambda\right\}_{\lambda\in\P^1(\C)}$, it fixes the value of the parameter $\lambda\in\P^1(\C)$ by \eqref{eq:lambda}, uniquely.
Note that the only point $P^\prime_\infty$ at infinity on $\overline\gamma_\lambda$ is in $\overline{\mathcal{B}}$.
Hence, for a positive integer $t>0$, we assume that the point $(z^s,w^s)$ is on the affine curve $\gamma_\lambda$ for $0\leq s\leq t$.
We moreover assume $(z^s,w^s)\not\in\overline{\mathcal{B}}$ for $0\leq s\leq t$ by virtue of the birationality of the map $\psi_{\rm vh}$.
We then have $f(z^t,w^t)=0$, or equivalently have
\begin{align*}
\lambda
=
-\frac{(w^t+1)^2+(z^t)^4}{(z^t)^2w^t}.
\end{align*}

First we consider the horizontal flip $\psi_{\rm h}:(z^t,w^t)\mapsto(z^{t+1},w^t)$.
By means of the map $\psi_{\rm h}$, the point $(z^t,w^t)\in\gamma_\lambda$ is mapped into the point $(z^{t+1},w^t)$.
We then compute
\begin{align*}
f(z^{t+1},w^t)
&=
(w^t+1)^2-\frac{(w^t+1)^2+(z^t)^4}{(z^t)^2} (z^{t+1})^2+(z^{t+1})^4\\
&=
(w^t+1)^2-(z^t)^2\left(\frac{w^t+1}{z^t}\right)^2
=0.
\end{align*}
Thus, the point $(z^{t+1},w^t)$ is also on the curve $\gamma_\lambda$.

Next we consider the vertical flip $\psi_{\rm v}:(z^{t+1},w^t)\mapsto(z^{t+1},w^{t+1})$.
We compute
\begin{align*}
f(z^{t+1},w^{t+1})
&=
(w^{t+1}+1)^2-\frac{(w^t+1)^2+(z^{t+1})^4}{w^t}w^{t+1}+(z^{t+1})^4\\
&=
\frac{w^{t+1}-w^t}{w^t}\left\{w^tw^{t+1}-1-(z^{t+1})^4\right\}
=
0,
\end{align*}
where we use $w^tw^{t+1}=(z^{t+1})^4+1$ and
\begin{align*}
\lambda
=
-\frac{(w^t+1)^2+(z^t)^4}{(z^t)^2w^t}
=
-\frac{(w^t+1)^2+(z^{t+1})^4}{(z^{t+1})^2w^t}.
\end{align*}
It follows that the point $(z^{t+1},w^{t+1})$ is also on $\gamma_\lambda$.
Induction on $t$ completes the proof.
\qed

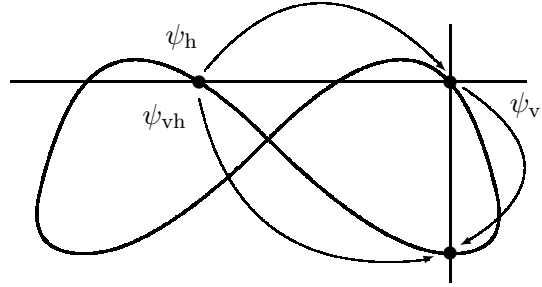
\begin{figure}[htb]
\centering
\unitlength=.03in
\def\arraystretch{1.0}
\begin{picture}(100,50)(-5,-2)
\qbezier(29,32)(48,55)(70,33)
\put(68,35){\vector(1,-1){3}}
\qbezier(74,29)(95,15)(75,2.3)
\put(76.5,3){\vector(-2,-1){3}}
\qbezier(28,27)(35,-5)(68,-1)
\put(65,-1.8){\vector(3,1){4}}
\thicklines
\qbezier(5,25)(15,45)(40,20)
\qbezier(5,25)(2,19)(0,10)
\qbezier(0,10)(-2,0)(8,0)
\qbezier(8,0)(20,0)(40,20)
\qbezier(40,20)(65,45)(75,25)
\qbezier(75,25)(78,19)(80,10)
\qbezier(80,10)(82,0)(72,0)
\qbezier(72,0)(60,0)(40,20)
\put(-5,30){\line(1,0){90}}
\put(71.8,40){\line(0,-1){45}}
\put(71.8,30){\circle*{2.5}}
\put(28,30){\circle*{2.5}}
\put(71.8,.2){\circle*{2.5}}
\put(25,38){\makebox(0,0){$\psi_{\rm h}$}}
\put(85,26){\makebox(0,0){$\psi_{\rm v}$}}
\put(22,24){\makebox(0,0){$\psi_{\rm vh}$}}
\end{picture}
\caption{
The horizontal flip $\psi_{\rm h}$, the vertical flip $\psi_{\rm v}$ and their composition $\psi_{\rm vh}=\psi_{\rm v}\circ\psi_{\rm h}$ on the singular curve $\gamma_\lambda$.
These maps are arising from the seed mutations $\mu_1$, $\mu_2$ and $\mu_2\circ\mu_1$ of type $A^{(2)}_2$, respectively.
}
\label{fig:Flips}
\end{figure}

\begin{remark}
The birational map $\psi_{\rm vh}$ is a map of QRT type \cite{Tsuda04}, {\it viz}, $\psi_{\rm vh}$ is the composition $\psi_{\rm v}\circ\psi_{\rm h}$ of the horizontal flip $\psi_{\rm h}:(z^t,w^t)\mapsto(z^{t+1},w^t)$ and the vertical flip $\psi_{\rm v}:(z^{t+1},w^t)\mapsto(z^{t+1},w^{t+1})$ (see proof of theorem \ref{thm:A22ds} and figure \ref{fig:Flips}).
This is the reason why we denote the map by $\psi_{\rm vh}$.
Moreover, the horizontal flip $\psi_{\rm h}$ and the vertical flip $\psi_{\rm v}$ are naturally arising from the seed mutations $\mu_1$ and $\mu_2$ of type $A^{(2)}_2$, respectively.
\end{remark}

\begin{remark}
We can derive the invariant curve $\overline\gamma_\lambda$ of the map $\psi_{\rm vh}$ with the aid of tropical geometry (see Appendix \ref{sec:App}).
\end{remark}

\subsection{Resolution of singularity}
Now we consider resolution of the singularity of $\overline\gamma_\lambda$ at $\mathfrak{p}$, which enables us to linearize the birational map $\psi_{\rm vh}$ of type $A^{(2)}_2$.
Since $\mathfrak{p}$ is an ordinary double point, we can resolve the singularity by blowing-up $\overline\gamma_\lambda$ at $\mathfrak{p}$ once.
Moreover, since $\mathfrak{p}$ is the base point of the pencil $\left\{\overline\gamma_\lambda\right\}_{\lambda\in\P^1(\C)}$, the singularity $\mathfrak{p}$ of the curves in the pencil are resolved by the blowing-up, all at once.

Let $\mathcal{U}\simeq\A^2\subset\P^2(\C)$ be an affine plane containing $\mathfrak{p}$.
Note that $\overline\gamma_\lambda\cap\mathcal{U}=\gamma_\lambda$.
Then the blowing-up $\widetilde U=\widetilde{\mathcal{U}_0}\cup\widetilde{\mathcal{U}_1}\subset\P^1(\C)\times\A^2$ of $\mathcal{U}$ at $\mathfrak{p}$ is given by
\begin{align*}
&\widetilde U
:=
\left\{
\left((a_0:a_1),(b,c)\right)\ |\
ba_1=(c+1)a_0
\right\},
\\
&\widetilde{\mathcal{U}_i}
:=
\left\{
\left((a_0:a_1),(b,c)\right)\in\widetilde U\ |\
a_i\neq0
\right\}\simeq\A^2
\quad
(i=0,1).
\end{align*}

\begin{figure}[htb]
\centering
\includegraphics[scale=.75]{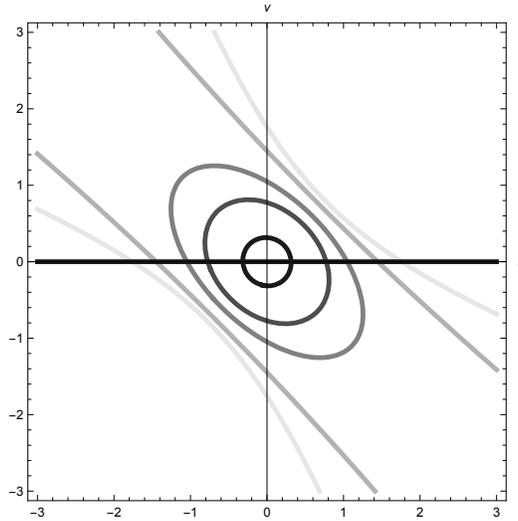}
\caption{The exceptional curve $v=0$ and the conics $\widetilde{\gamma}_\lambda$ for $\lambda=0.1$, $0.6$, $1.1$, $2.1$ and $3.1$.
A curve with smaller $\lambda$ is colored darker.
Each curve in the pencil $\left\{\widetilde{\gamma}_\lambda\right\}_{\lambda\in\P^1(\C)}$ intersects the exceptional curve at the points $\left(\pm\sqrt{\lambda},0\right)$.}
\label{fig:pencilresolved}
\end{figure}

On the affine plane $\widetilde{\mathcal{U}_0}$, the projection $\pi:\widetilde U\to\mathcal{U}$ is defined to be
\begin{align*}
\pi:(u,v)\mapsto (z,w)
=(v,uv-1).
\end{align*}
The total transform of $\gamma_\lambda$ is $v^2\widetilde {f_0}(u,v)=0$, where we define
\begin{align*}
&\widetilde  {f_0}(u,v)
:=
u^2+\lambda(uv-1)+v^2.
\end{align*}
We denote the conic given by $\widetilde {f_0}(u,v)=0$ by $\widetilde{\gamma}_\lambda$.
The exceptional curve is given by $v=0$.
These curves, $\widetilde{\gamma}_\lambda$ and $v=0$, intersect at the two points (see figure \ref{fig:pencilresolved})
\begin{align*}
(u,v)=\left(\pm\sqrt{\lambda},0\right).
\end{align*}

Let us transform the birational map $\psi_{\rm vh}$ of type $A^{(2)}_2$ on the singular curve $\gamma_\lambda$ into the one $\widetilde\psi_{\rm dv}=\pi^{-1}\circ\psi_{\rm vh}\circ\pi$ on the conic $\widetilde{\gamma}_\lambda$ by means of the blowing-up $\pi$:
\begin{align*}
&\begin{CD}
\gamma_\lambda @<\pi<< \widetilde\gamma_\lambda
\\
@A \psi_{\rm vh} AA @AA \widetilde\psi_{\rm dv} A
\\
\gamma_\lambda @<\pi<< \widetilde\gamma_\lambda.
\end{CD}
\end{align*}
We refer to the map $\widetilde\psi_{\rm dv}$ on the conic $\widetilde{\gamma}_\lambda$ and to the dynamical system governed by it (see proposition \ref{eq:brmoc} below) as of type $A^{(2)}_2$ as well as the map $\psi_{\rm vh}$ on the singular curve $\gamma_\lambda$.

\begin{proposition}
\label{eq:brmoc}
The birational map $\widetilde\psi_{\rm dv}=\pi^{-1}\circ\psi_{\rm vh}\circ\pi:$
\begin{align}
(u^t,v^t)\mapsto(u^{t+1},v^{t+1})
=
\left(
\frac{(u^t)^3+v^t}{u^tv^t-1},
u^t
\right)
\label{eq:tildepsi}
\end{align}
induces a dynamical system on the conic $\widetilde{\gamma}_\lambda$ from which 4 points are removed:
\begin{align*}
\begin{cases}
\widetilde{\gamma}_\lambda
-
\widetilde{\mathcal{B}},
\\
\widetilde{\gamma}_\lambda
=
\left(
\widetilde  {f_0}(u,v)=0
\right)
=
\left(
u^2+\lambda(uv-1)+v^2=0
\right),
\\[5pt]
\widetilde{\mathcal{B}}
:=
\left\{
(\zeta_8^{8-i},\zeta_8^i)\ |\ 
i=1,3,5,7
\right\}.
\end{cases}
\end{align*}
Moreover, the map $\widetilde\psi_{\rm dv}$ is the composition $\widetilde\psi_{\rm d}\circ\widetilde\psi_{\rm v}$ of the vertical flip $\widetilde\psi_{\rm v}$ and the diagonal flip $\widetilde\psi_{\rm d}$:
\begin{align}
&\widetilde\psi_{\rm v}:
(u^t,v^t)\mapsto\left(u^t,\frac{(u^t)^3+v^t}{u^tv^t-1}\right),
\label{eq:psiv}
\\
&\widetilde\psi_{\rm d}:
(u^t,v^t)\mapsto(v^t,u^t)
\label{eq:psid}
\end{align}
on the conic $\widetilde{\gamma}_\lambda$.
\end{proposition}

(Proof)\quad
First note that the base points $[\zeta_8^i:0:1]$ ($i=1,3,5,7$) of the pencil $\left\{\gamma_\lambda\right\}_{\lambda\in\C}$ are mapped into the base points $(u,v)=(\zeta_8^{8-i},\zeta_8^i)$ ($i=1,3,5,7$) of the pencil $\left\{\widetilde{\gamma}_\lambda\right\}_{\lambda\in\C}$ by $\pi^{-1}$, respectively.
There exists no other base points of $\left\{\widetilde{\gamma}_\lambda\right\}_{\lambda\in\C}$.
The base point $\mathfrak{p}$ of $\left\{\gamma_\lambda\right\}_{\lambda\in\C}$ is transformed into the exceptional curve $v=0$, which intersects $\widetilde{\gamma}_\lambda$ at $\left(\pm\sqrt{\lambda},0\right)$.
Since the inverse $\pi^{-1}$ is not uniquely defined at these points, we extend $\widetilde\psi_{\rm dv}$ at $\left(\pm\sqrt{\lambda},0\right)$ by using \eqref{eq:tildepsi}.
Therefore, we assume that the initial point $(u^0,v^0)$ satisfies $(u^0,v^0)\not\in\widetilde{\mathcal{B}}$.
Then the value of $\lambda$ is uniquely fixed
\begin{align*}
\lambda
=
-\frac{(u^0)^2+(v^0)^2}{u^0v^0-1}.
\end{align*}

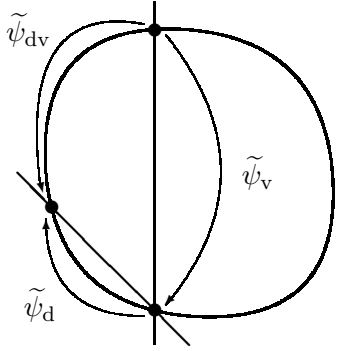
\begin{figure}[htb]
\centering
\unitlength=.03in
\def\arraystretch{1.0}
\begin{picture}(70,70)(-33,-30)
\qbezier(-4,24)(15,0)(-4,-23)
\put(-1.3,-20){\vector(-1,-1){3}}
\qbezier(-8,-25)(-25,-25)(-25,-8)
\put(-25,-9){\vector(0,1){1}}
\qbezier(-8,26)(-30,30)(-26,-2)
\put(-26.5,0.5){\vector(1,-4){1}}
\thicklines
\qbezier(-25,0)(-27,27)(0,25)
\qbezier(0,25)(23,23)(25,0)
\qbezier(25,0)(27,-27)(0,-25)
\qbezier(0,-25)(-23,-23)(-25,0)
\put(0,-30){\line(-1,1){30}}
\put(-6,30){\line(0,-1){60}}
\put(-6,-24.){\circle*{2.5}}
\put(-6,25){\circle*{2.5}}
\put(-24,-6){\circle*{2.5}}
\put(-26,-23){\makebox(0,0){$\widetilde\psi_{\rm d}$}}
\put(12,0){\makebox(0,0){$\widetilde\psi_{\rm v}$}}
\put(-28,25){\makebox(0,0){$\widetilde\psi_{\rm dv}$}}
\end{picture}
\caption{
The vertical flip $\widetilde\psi_{\rm d}$, the diagonal flip $\widetilde\psi_{\rm d}$ and their composition $\widetilde\psi_{\rm dv}=\widetilde\psi_{\rm d}\circ\widetilde\psi_{\rm v}$ on the conic $\widetilde\gamma_\lambda$.
The birational map $\widetilde\psi_{\rm dv}=\pi^{-1}\circ\psi_{\rm vh}\circ\pi$ is conjugate to the birational map $\psi_{\rm vh}$ with respect to $\pi$, both of which are arising from the seed mutations of type $A^{(2)}_2$.
}
\label{fig:diagonalflip}
\end{figure}

Let $t>0$ be a positive integer.
Assume that the point $(u^s,v^s)$ is on $\widetilde{\gamma}_\lambda$ for $0\leq s\leq t$.
We then have $\widetilde{f}(u^t,v^t)=0$, or equivalently have
\begin{align*}
\lambda
=
-\frac{(u^t)^2+(v^t)^2}{u^tv^t-1}.
\end{align*}
Consider the vertical flip $\widetilde\psi_{\rm v}$ \eqref{eq:psiv}.
We show that the point $\left(u^t,\frac{(u^t)^3+v^t}{u^tv^t-1}\right)$ is also on $\widetilde{\gamma}_\lambda$:
\begin{align*}
\widetilde{f}_0\left(u^t,\frac{(u^t)^3+v^t}{u^tv^t-1}\right)
&=
(u^t)^2-\frac{(u^t)^4(v^t)^2+(u^t)^2-2(u^t)^3v^t}{(u^tv^t-1)^2}
=0.
\end{align*}
Since $\widetilde{\gamma}_\lambda$ is symmetric with respect to the line $v=u$, the diagonal flip $\widetilde\psi_{\rm d}$ \eqref{eq:psid} is obviously a map on $\widetilde{\gamma}_\lambda$.
We then conclude that the composition $\widetilde\psi_{\rm dv}=\widetilde\psi_{\rm d}\circ\widetilde\psi_{\rm v}$ is a map on $\widetilde{\gamma}_\lambda-\widetilde{\mathcal{B}}$ (see figure \ref{fig:diagonalflip}).
Induction on $t$ completes the proof.
\qed

We thus transform the birational map $\psi_{\rm vh}$ on the singular quartic curve $\gamma_\lambda$ arising from the seed mutations of type $A^{(2)}_2$ into the birational map $\widetilde\psi_{\rm dv}$ on the non-singular conic $\widetilde\gamma_\lambda$ by using the blowing-up of the plane. 

In the subsequent section, we show that the above geometric approach via the invariant curve and its blowing-up enables us to linearize the seed mutations of type $A^{(2)}_2$.
We give a linearization of the map $\widetilde\psi_{\rm dv}$ explicitly with the aid of the linearizable QRT map arising from the seed mutations of type $A^{(1)}_1$.
It is straightforward to construct the set of generators of the cluster algebra whose seed mutations are linearized.

\section{Linearization of seed mutations and generators of cluster algebras}
\label{sec:A11A22}
Now we linearize the birational map $\widetilde\psi_{\rm dv}$ on the conic $\widetilde\gamma_\lambda$ arising from the seed mutations of type $A^{(2)}_2$.
The conic $\widetilde\gamma_\lambda$ is essentially the same as the invariant curve of the linearizable QRT map $\phi_{\rm vh}$ arising from the seed mutations of type $A^{(1)}_1$ (see \eqref{eq:QRTA11} below).

\subsection{Seed mutations of type $\boldsymbol{A^{(1)}_1}$}
\label{sec:MA11}
In the preceding paper \cite{Nobe16}, we investigated the birational map $\phi_{\rm vh}:\P^2(\C)\to\P^2(\C);$
\begin{align}
&(z^t,w^t)\mapsto(z^{t+1},w^{t+1})
=
\left(
\frac{(w^t)^2+1}{z^t},
\frac{(z^{t+1})^2+1}{w^t}
\right)
\label{eq:QRTA11}
\end{align}
which is  arising from the seed mutations of type $A^{(1)}_1$ in the following manner.

Let $\mathcal{A}$ be the cluster algebra whose initial seed $(\bx_0,\by_0,B_0)$ is given by
 \begin{align}
\bx_0=\left(x_1,x_2\right),\quad
\by_0=\left(y_1,y_2\right),\quad
B_0=\left(\begin{matrix}0&-2\\2&0\\\end{matrix}\right).
\label{eq:ISA11}
\end{align}
We consider the cluster pattern $\T_2\ni t_m\mapsto\Sigma_m=\left(\bx_m,\by_m,B_m\right)$ in figure \ref{fig:binarytree}.
We associate the variables $z^t$ and $w^t$ with the seeds of $\mathcal{A}$ as
\begin{align}
z^t
=
\frac{x_{1;2t}}{\sqrt{y_{2;2t}}},
\quad
w^t
=
\sqrt{y_{1;2t}}x_{2;2t}
\quad
(t\geq0),
\label{eq:zwxyA11}
\end{align}
where we define $x_{1;2t}$, $x_{2;2t}$, $y_{1;2t}$ and $y_{2;2t}$ as in \eqref{eq:mu1mu2x} and \eqref{eq:mu1mu2y}.
All the Cartan counterparts $A(B_m)$ of the exchange matrices $B_m$ ($m\in\Z$, see \eqref{eq:mu1mu2B}) are of type $A^{(1)}_1$.
Hence, we refer to the cluster algebra and to the seed mutations as of type $A^{(1)}_1$.
The semifield $\P$ is arbitrarily chosen.
Then the exchange relations \eqref{eq:mutcoef} and \eqref{eq:mutcv} induce the birational map $\phi_{\rm vh}$ \cite{Nobe16}.
We also refer to the birational map $\phi_{\rm vh}$ and to the dynamical system governed by it as of type $A^{(1)}_1$.

Since we easily see that the birational map $\phi_{\rm vh}$ is a member of the family of QRT maps, the invariant curve $\overline\delta_\nu$ of $\phi_{\rm vh}$ is automatically obtained as follows \cite{QRT89}
\begin{align*}
\begin{cases}
\overline\delta_\nu
:=
\delta_\nu
\cup
\left\{
Q_\infty^+,
Q_\infty^-
\right\},
\\
\delta_\nu
:=
\left(
g(z,w)
:=
z^2+\nu zw+w^2+1
=
0
\right),
\end{cases}
\end{align*}
where $\nu$ is the conserved quantity and the points $Q_\infty^+$ and $Q_\infty^-$ at infinity are given by
\begin{align*}
Q_\infty^+
=
[-\nu+\nu^\prime :2:0],
\quad
Q_\infty^-
=
[-\nu-\nu^\prime:2:0]
\end{align*}
in the homogeneous coordinate $(z,w)\mapsto[z:w:1]$.
Here we put $\nu^\prime :=\sqrt{\nu^2-4}$.
The base points of the pencil $\left\{\overline\delta_\nu\right\}_{\nu\in\P^1(\C)}$ are the following four points:
\begin{align}
\left[0:\pm\sqrt{-1}:1\right],
\quad
\left[\pm\sqrt{-1}:0:1\right].
\label{eq:BPA11}
\end{align}

\begin{remark}\label{rem:ellipticcurveconic}
In general, the invariant curve of a QRT map is an elliptic curve.
The invariant curve $\overline\delta_\nu$ of the birational map $\phi_{\rm vh}$, however, is not an elliptic curve but a conic.
It should be noted that the family of QRT maps contains such degenerated maps on conics by definition \cite{QRT89}.
Therefore, the map $\phi_{\rm vh}$ is nothing but the QRT map.
The maps on conics also have the flipping structure, \textit{viz}, they are decomposed into horizontal flips and vertical flips.
\end{remark}


It is well known that the QRT map $\phi_{\rm vh}$ is linearizable \cite{RGSM11}.
Actually, with the aid of the conserved quantity, we have
\begin{align*}
z^{t+1}
&=
\frac{(w^t)^2+1}{z^t}
=
\frac{-(z^t)^2-\nu z^tw^t}{z^t}
=
-z^t-\nu w^t,
\\
w^{t+1}
&=
\frac{(z^{t+1})^2+1}{w^t}
=
\frac{-(w^{t+1})^2-\nu z^{t+1}w^{t+1}}{w^t}
=
\frac{\left(-w^{t+1}+\nu z^t+\nu^2 w^t\right)w^{t+1}}{w^t}.
\end{align*}
Since $w^{t+1}\not\equiv0$, we obtain the linearization of the QRT map $\phi_{\rm vh}$:
\begin{align}
(z^t,w^t)
\mapsto
(z^{t+1},w^{t+1})
=
\left(
-z^t-\nu w^t,
\nu z^t+(\nu^2-1)w^t
\right).
\label{eq:lmA11}
\end{align}

\subsection{Linearization}
\label{sec:comm}
Now introduce the linear map $\omega:\P^2(\C)\to\P^2(\C)$ depending on the parameter $\lambda\in\P^1(\C)$:
\begin{align*}
\omega:(u,v)\mapsto(z,w)=\left(-\frac{u}{\sqrt{-\lambda}}, -\frac{v}{\sqrt{-\lambda}}\right).
\end{align*}
The conic $\widetilde{\gamma}_\lambda$, the invariant curve of the birational map $\widetilde\psi_{\rm dv}$ of type $A^{(2)}_2$, is mapped into the conic $\delta_\lambda$, the invariant curve of the linearizable QRT map $\phi_{\rm vh}$, by means of $\omega$.
\begin{proposition}\label{prop:gammadelta}
If $(u,v)\in\widetilde{\gamma}_\lambda$ then $\omega(u,v)\in\delta_\lambda$, and vice versa.
\end{proposition}

(Proof)\quad
Let $(z,w)$ be $\omega(u,v)$.
Then we have
\begin{align*}
u
=
-\sqrt{-\lambda}z,
\quad
v=
-\sqrt{-\lambda}w.
\end{align*}
We compute
\begin{align*}
u^2+\lambda(uv-1)+v^2
=
-\lambda\left(z^2+\lambda z w+ w^2+1\right).
\end{align*}
The statement is obvious already.
\qed

We then obtain the birational map $\phi_{\rm dv}:=\omega\circ\widetilde\psi_{\rm dv}\circ\omega^{-1}:$
\begin{align}
&(z^t, w^t)
\mapsto
(z^{t+1}, w^{t+1})
=
\left(\frac{(z^t)^2+1}{ w^t}, z^t\right)
\label{eq:A22ondelta}
\end{align}
on $\delta_\lambda$ conjugate to the map $\widetilde\psi_{\rm dv}$ on $\widetilde{\gamma}_\lambda$ with respect to the linear map $\omega$:
\begin{align*}
\begin{CD}
\gamma_\lambda @<\pi<< \widetilde\gamma_\lambda @> \omega >>\delta_\lambda @> \phi_{\rm vh} >>\delta_\lambda
\\
@A \psi_{\rm vh} AA @A \widetilde\psi_{\rm dv} AA @A \phi_{\rm dv} AA @VV \phi_{\rm dv} V
\\
\gamma_\lambda @<<\pi< \widetilde\gamma_\lambda @>> \omega >\delta_\lambda @<< \phi_{\rm vh} <\delta_\lambda.
\end{CD}
\end{align*}

The birational map $\phi_{\rm dv}$ thus obtained is referred to as of type $A^{(2)}_2$ as well as the ones $\psi_{\rm vh}$ and $\widetilde\psi_{\rm dv}$.
Moreover, the map $\phi_{\rm dv}$ has a flipping structure similar to QRT maps and commutes with the linearizable QRT map $\phi_{\rm vh}$ on $\delta_\lambda$.
\begin{thm}\label{thm:A11A22}
Let $\phi_{\rm h}$, $\phi_{\rm v}$ and $\phi_{\rm d}$ be the horizontal flip, the vertical flip and the diagonal flip on the conic $\delta_\lambda$, respectively:
\begin{align*}
&\phi_{\rm h}:(z^t, w^t)\mapsto
\left(
\frac{( w^t)^2+1}{ z^t}, 
w^t
\right),
\\
&\phi_{\rm v}:(z^t, w^t)\mapsto
\left(
z^t, 
\frac{(z^t)^2+1}{ w^t}
\right),
\\
&\phi_{\rm d}:(z^t, w^t)\mapsto( w^t, z^t).
\end{align*}
Then we have the following decomposition of the birational maps $\phi_{\rm vh}$ of type $A^{(1)}_1$ (see \eqref{eq:QRTA11}) and $\phi_{\rm dv}$ of type $A^{(2)}_2$ (see \eqref{eq:A22ondelta}):
\begin{align*}
\phi_{\rm vh}=\phi_{\rm v}\circ\phi_{\rm h}
\quad
\mbox{and}
\quad
\phi_{\rm dv}=\phi_{\rm d}\circ\phi_{\rm v}.
\end{align*}
Moreover, these birational maps are commutative:
\begin{align*}
\phi_{\rm vh}\circ\phi_{\rm dv}=\phi_{\rm dv}\circ\phi_{\rm vh}.
\end{align*}
\end{thm}

(Proof)\quad
It is clear that $\phi_{\rm vh}=\phi_{\rm v}\circ\phi_{\rm h}$ holds since the birational map $\phi_{\rm vh}$ is a QRT map.

Fix a certain $t>0$.
Suppose that $(u^s,v^s)\in\widetilde{\gamma}_\lambda$ for $0\leq s\leq t$.
We then have
\begin{align*}
\widetilde{f_0}(u^t,v^t)
=
(u^t)^2+\lambda\left(u^tv^t-1\right)+(v^t)^2
=
0.
\end{align*}
If $\omega(u^t,v^t)=(z^t, w^t)$ then we also have
\begin{align*}
g(z^t, w^t)
=
(z^t)^2+\lambda z^t w^t+( w^t)^2+1
=
0
\end{align*}
by proposition \ref{prop:gammadelta}.

Now we assume that $(u^t,v^t)$ and $(u^{t+1},v^{t+1})$ satisfy $(u^{t+1},v^{t+1})=\widetilde\psi_{\rm dv}(u^t,v^t)$, where $\widetilde\psi_{\rm dv}$ is the birational map of type $A^{(2)}_2$ (see \eqref{eq:tildepsi}).
By substituting $z^t=-u^t/\sqrt{-\lambda}$ and $w^t=-v^t/\sqrt{-\lambda}$ into \eqref{eq:tildepsi}, we obtain
\begin{align*}
-\sqrt{-\lambda} z^{t+1}
&=
\frac{\sqrt{-\lambda}^3(z^t)^3-\sqrt{-\lambda} w^t}{-\lambda z^t w^t-1}.
\end{align*}
Noting that $\lambda$ is the conserved quantity:
\begin{align*}
\lambda
=
-\frac{(z^t)^2+( w^t)^2+1}{ z^t w^t},
\end{align*}
we have
\begin{align*}
 z^{t+1}
&=
\frac{\lambda(z^t)^3- w^t}{\lambda z^t w^t+1}
=
\frac{(z^t)^2+1}{ w^t}.
\end{align*}
We similarly have
\begin{align*}
 w^{t+1}
&=
 z^t.
\end{align*}
Therefore, the birational map $\phi_{\rm dv}$ \eqref{eq:A22ondelta} is the composition $\phi_{\rm d}\circ\phi_{\rm v}$ of the vertical flip $\phi_{\rm v}$ and the diagonal flip $\phi_{\rm d}$.

Next we show the commutativity.
Noting that the vertical flip $\phi_{\rm v}$ is an involution, the condition $\phi_{\rm vh}\circ\phi_{\rm dv}=\phi_{\rm dv}\circ\phi_{\rm vh}$ to be confirmed reduces to
\begin{align}
\phi_{\rm v}\circ\phi_{\rm h}\circ\phi_{\rm d}
=
\phi_{\rm d}\circ\phi_{\rm h}\circ\phi_{\rm v}.
\label{eq:tobeconfirmed}
\end{align}
We then compute the LHS:
\begin{align*}
(z, w)
&\overset{\phi_{\rm d}}{\mapsto}
\left( w, z\right)
\overset{\phi_{\rm h}}{\mapsto}
\left(\frac{ z^2+1}{ w}, z\right)
\overset{\phi_{\rm v}}{\mapsto}
\left(\frac{ z^2+1}{ w},\frac{\left(\frac{ z^2+1}{ w}\right)^2+1}{ z}\right)
\end{align*}
and the RHS:
\begin{align*}
 (z, w)
&\overset{\phi_{\rm v}}{\mapsto}
\left(z,\frac{ z^2+1}{ w}\right)
\overset{\phi_{\rm h}}{\mapsto}
\left(\frac{\left(\frac{ z^2+1}{ w}\right)^2+1}{ z},\frac{ z^2+1}{ w}\right)
\overset{\phi_{\rm d}}{\mapsto}
\left(\frac{ z^2+1}{ w},\frac{\left(\frac{ z^2+1}{ w}\right)^2+1}{ z}\right).
\end{align*}
Thus, \eqref{eq:tobeconfirmed} holds for any $(z, w)\in\delta_\nu$ other than the base points \eqref{eq:BPA11}.
\qed

From theorem \ref{thm:A11A22}, a corollary concerning linearization of the birational map $\phi_{\rm dv}$ of type $A^{(2)}_2$ follows immediately.
\begin{corollary}
\label{cor:linearizationA22}
The birational map $\phi_{\rm dv}$ of type $A^{(2)}_2$ on the conic $\delta_\lambda$, which commutes with the linearizable QRT map $\phi_{\rm vh}$ of type $A^{(1)}_1$ on $\delta_\lambda$, is also linearizable:
\begin{align}
\phi_{\rm dv}:\ (z^t, w^t)
\mapsto
(z^{t+1}, w^{t+1})
=
\left(-\lambda z^t- w^t, z^t\right),
\label{eq:lmA22}
\end{align}
where $\lambda$ is given by
\begin{align*}
\lambda
=
-\frac{(z^0)^2+( w^0)^2+1}{ z^0 w^0}.
\end{align*}
\end{corollary}

(Proof)\quad
This is a direct consequence of theorem \ref{thm:A11A22} and the linearization \eqref{eq:lmA11} of the map $\phi_{\rm vh}$.
Note that the linearization of the map $\widetilde\psi_{\rm dv}$ on the conic $\widetilde\gamma_\lambda$ is also given by \eqref{eq:lmA22} since $\widetilde\psi_{\rm dv}$ and $\phi_{\rm dv}$ are transformed into each other by the linear map $\omega$.
\qed

Now we summarize all birational maps arising from the seed mutations of types $A^{(1)}_1$ and $A^{(2)}_2$ in table \ref{tab:bmiv}.
The invariant curves are also summarized in the table.
\begin{table*}[htbp]\centering
\caption{
Properties of the birational maps and their invariant curves arising from the seed mutations of types $A^{(1)}_1$ and $A^{(2)}_2$.
}
\label{tab:bmiv}
{\renewcommand\arraystretch{1.}
\begin{ruledtabular}
\begin{tabular}{lllllllll}
Type&Map&Eq.&Linearizability\footnotemark[1]&\multicolumn{2}{l}{Relation among maps}&Curve&Degree\footnotemark[2]&Singularity\footnotemark[3]
\\
\hline
$A^{(2)}_2$
&
$\psi_{\rm vh}$
&
\eqref{eq:bmA22}
&
No
&
\multicolumn{1}{c}{Conjugate}
&
&
$\gamma_\lambda$
&
4
&
Yes
\\[3pt]
$A^{(2)}_2$
&
$\widetilde\psi_{\rm dv}$
&
\eqref{eq:tildepsi}
&
Yes
\quad
\eqref{eq:lmA22}
&
\multicolumn{1}{c}{with}
&
&
$\widetilde\gamma_\lambda$
&
2
&
No
\\[3pt]
$A^{(2)}_2$
&
$\phi_{\rm dv}$
&\eqref{eq:A22ondelta}
&
Yes
\quad
\eqref{eq:lmA22}
&
\multicolumn{1}{c}{each other}
&
\multicolumn{1}{c}{Commutative}
&
\multirow{2}{*}{$\delta_\lambda$}
&
\multirow{2}{*}{2}
&
\multirow{2}{*}{No}
\\[3pt]
$A^{(1)}_1$
&
$\phi_{\rm vh}$
&
\eqref{eq:QRTA11}
&
Yes
\quad
\eqref{eq:lmA11}
&
&
\multicolumn{1}{c}{with each other}
&
&
&
\\
\end{tabular}
\footnotetext[1]{Whether the map is linearizable. The equation giving the linearization is also presented.}
\footnotetext[2]{The degree of the invariant curve.}
\footnotetext[3]{Whether the invariant curve is singular.}
\end{ruledtabular}
}
\end{table*}

\subsection{General solutions}
\label{sec:GS}

From the linearization \eqref{eq:lmA11} of the QRT map $\phi_{\rm vh}$ of type $A^{(1)}_1$, it immediately follows the general solution to the initial value problem of the dynamical system governed by the map.
\begin{thm}
Let $(z^0,w^0)$ be a point on $\P^2(\C)$ other than the base points \eqref{eq:BPA11} of the pencil $\left\{\overline\delta_\nu\right\}_{\nu\in\P^1(\C)}$.
Then the general solution to the dynamical system
\begin{align*}
(z^{t+1},w^{t+1})
=
\phi_{\rm vh}(z^t,w^t)
=
\left(
\frac{\left(w^t\right)^2+1}{z^t},
\frac{\left(z^{t+1}\right)^2+1}{w^t}
\right)
\end{align*}
governed by the birational map $\phi_{\rm vh}$ arising from the seed mutations of type $A^{(1)}_1$ is given by
\begin{align*}
z^t
=
z^0\ch(Mt)-\frac{\nu z^0+2w^0}{\nu^\prime }\sh(Mt),
\quad
w^t
=
w^0\ch(Mt)+\frac{\nu w^0+2z^0}{\nu^\prime }\sh(Mt),
\end{align*}
where we put
\begin{align*}
\begin{cases}
\DIS
M
=\log\left(\nu^2-2+\nu\nu^\prime \right)-\log2,
\\
\DIS
\nu
=
-\frac{(z^0)^2+(w^0)^2+1}{z^0w^0},\\
\nu^\prime 
=
\sqrt{\nu^2-4}\\
\end{cases}
\end{align*}
and the hyperbolic functions $\cosh\theta$ and $\sinh\theta$ are abbreviated to $\ch\theta$ and $\sh\theta$, respectively.
\end{thm}

(Proof)\quad
We compute \eqref{eq:lmA11} as follows
\begin{align*}
\left(
\begin{matrix}
z^{t}\\
w^{t}\\
\end{matrix}
\right)
&=
\left(
\begin{matrix}
-1&-\nu\\
\nu&\nu^2-1\\
\end{matrix}
\right)
\left(
\begin{matrix}
z^{t-1}\\
w^{t-1}\\
\end{matrix}
\right)
\\
&=
\left(
\begin{matrix}
\DIS\ch(Mt)-\frac{\nu}{\nu^\prime }\sh(Mt)
&
\DIS-\frac{2}{\nu^\prime }\sh(Mt)\\
\DIS\frac{2}{\nu^\prime }\sh(Mt)
&
\DIS\ch(Mt)+\frac{\nu}{\nu^\prime }\sh(Mt)\\
\end{matrix}
\right)
\left(
\begin{matrix}
z^{0}\\
w^{0}\\
\end{matrix}
\right).
\end{align*}
This clearly gives the general solution.
\qed

The general solution to the dynamical system governed by the birational map $\psi_{\rm vh}$ of type $A^{(2)}_2$ is similarly obtained
\begin{thm}
\label{thm:GS}
Let $(z^0,w^0)$ be a point on $\P^2(\C)-\overline{\mathcal{B}}$ (see \eqref{eq:setofbasepoints}).
Then the general solution to the dynamical system
\begin{align*}
(z^{t+1},w^{t+1})
=
\psi_{\rm vh}(z^t,w^t)
=
\left(
\frac{w^t+1}{z^t},
\frac{\left(z^{t+1}\right)^4+1}{w^t}
\right)
\end{align*}
governed by the birational map $\psi_{\rm vh}$ arising from the seed mutations of type $A^{(2)}_2$ is given by
\begin{align*}
z^t
=
\rho(t),
\quad
w^t
=
\rho(t+1)\rho(t)-1,
\end{align*}
where we put
\begin{align}
\begin{cases}
\DIS
\rho(t)
=
\frac{2}{\lambda^\prime}\left(\frac{w^0+1}{z^0}\sh(\Lambda t)-z^0\sh(\Lambda (t-1))\right),
\\
\DIS
\Lambda
=\log\left(\lambda^\prime-\lambda\right)-\log2,
\\
\DIS
\lambda
=
-\frac{\left(w^0+1\right)^2+\left(z^0\right)^4}{\left(z^0\right)^2w^0},
\\
\lambda^\prime
=
\sqrt{\lambda^2-4}.
\end{cases}
\label{eq:rho}
\end{align}
\end{thm}

(Proof)\quad
First we give the general solution to the dynamical system $(x^{t+1}, y^{t+1})=\phi_{\rm dv}(x^t,y^t)$ governed by the birational map $\phi_{\rm dv}$ of type $A^{(2)}_2$ on the conic $\delta_\lambda$.
From the linearization of $\phi_{\rm dv}$ \eqref{eq:lmA22}, we have
\begin{align*}
\left(
\begin{matrix}
x^t
\\
y^t
\end{matrix}
\right)
&=
\left(
\begin{matrix}
-\lambda&-1
\\
1&0
\\
\end{matrix}
\right)^t
\left(
\begin{matrix}
x^{0}
\\
y^{0}
\end{matrix}
\right)
\\
&=
\frac{2}{\lambda^\prime}
\left(
\begin{matrix}
\sh(\Lambda (t+1))&-\sh(\Lambda t)
\\
\sh(\Lambda t)&-\sh(\Lambda (t-1))
\\
\end{matrix}
\right)
\left(
\begin{matrix}
x^{0}
\\
y^{0}
\end{matrix}
\right).
\end{align*}
Then the variable transformation $\pi\circ\omega^{-1}:(x^t,y^t)\mapsto(z^t,w^t)=(-\sqrt{-\lambda}y^t,-\lambda x^ty^t-1)$ leads to the conclusion. 
\qed

\subsection{Generators of cluster algebras}
\label{sec:GCA}
In order to construct the set of all cluster variables generating the cluster algebras of types $A^{(1)}_1$ and $A^{(2)}_2$, respectively, we give their coefficients explicitly. 
Let the semifield $\P$ of the coefficients be the tropical semifield $\left({\rm Trop}(y_1,y_2),\oplus,\cdot\right)$ \cite{FZ07} and fix it hereafter.

\begin{proposition}
The general terms of the coefficients $\by_{m}=(y_{1;m},y_{2;m})$ in the cluster algebra of type $A^{(2)}_2$, which start from the initial seed $\Sigma_0=\left(\bx_0,\by_0,B_0\right)$ (see \eqref{eq:initialseed}) and mutate along the cluster pattern in figure \ref{fig:binarytree}, are explicitly given by
\begin{align}
&\by_{2k}
=
\left(
y_1^{-2k+1}y_2^{-k+1},
y_1^{4k-4}y_2^{2k-3}
\right),
\quad
\by_{2k+1}
=
\left(
y_1^{2k-1}y_2^{k-1},
y_1^{-4k}y_2^{-2k+1}
\right)
\label{eq:coefficient}
\end{align}
for $k\geq1$.
\end{proposition}

(Proof)\quad
It is easy to check that \eqref{eq:coefficient} holds for $k=1$.

We assume that these are true for $k$.
We then have
\begin{align*}
\by_{2k+1}
\overset{\mu_2}{\longleftrightarrow}
\by_{2k+2}
&=
\left(
y_{1;2k+1}y_{2;2k+1}^0\left(y_{2;2k+1}\oplus1\right)^1,
y_{2;2k+1}^{-1}
\right)
\\
&=
\left(
y_1^{-2(k+1)+1}y_2^{-(k+1)+1},
y_1^{4(k+1)-4}y_2^{2(k+1)-3}
\right)
\end{align*}
and
\begin{align*}
\by_{2k+2}
\overset{\mu_1}{\longleftrightarrow}
\by_{2k+3}
&=
\left(
y_{1;2k+2}^{-1},
y_{2;2k+2}y_{1;2k+2}^0\left(y_{1;2k+2}\oplus1\right)^4
\right)
\\
&=
\left(
y_1^{2(k+1)-1}y_2^{(k+1)-1},
y_1^{-4(k+1)}y_2^{-2(k+1)+1}
\right),
\end{align*}
where we use the exchange relation \eqref{eq:mutcoef} for the coefficients and the fact for the exchange matrices:
\begin{align*}
B_m
=
\begin{cases}
B_0=\left(\begin{matrix}0&-4\\1&0\\\end{matrix}\right)&\mbox{$m$ even,}\\[20pt]
-B_0=\left(\begin{matrix}0&4\\-1&0\\\end{matrix}\right)&\mbox{$m$ odd.}\\
\end{cases}
\end{align*}
Induction on $k$ completes the proof.
\qed

Similarly, we obtain the following \cite{Nobe16}.
\begin{proposition}
The general terms of the coefficients $\by_{m}=(y_{1;m},y_{2;m})$ in the cluster algebra of type $A^{(1)}_1$, which start from the initial seed $\Sigma_0=\left(\bx_0,\by_0,B_0\right)$ (see \eqref{eq:ISA11}) and mutate along the cluster pattern in figure \ref{fig:binarytree}, are explicitly given by
\begin{align*}
&\by_{2k}
=
\left(
y_1^{-2k+1}y_2^{-2k+2},
y_1^{2k-2}y_2^{2k-3}
\right),
\quad
\by_{2k+1}
=
\left(
y_1^{2k-1}y_2^{2k-2},
y_1^{-2k}y_2^{-2k+1}
\right)
\end{align*}
for $k\geq1$.
\qed
\end{proposition}

With the aid of the relation \eqref{eq:zwxy}, we have
\begin{align*}
&\bx_{0}
=
\left(
y_2^{\frac{1}{4}}z^0,
y_1^{-1}w^{0}
\right),
\quad
\bx_{1}
=
\left(
y_{2}^{-\frac{1}{4}}z^{1},
y_1^{-1}w^{0}
\right)
\end{align*}
and
\begin{align*}
&\bx_{2k}
=
\left(
y_1^{k-1}y_2^\frac{2k-3}{4}z^k,
y_1^{2k-1}y_2^{k-1}w^{k}
\right),
\quad
\bx_{2k+1}
=
\left(
y_1^{k}y_2^\frac{2k-1}{4}z^{k+1},
y_1^{2k-1}y_2^{k-1}w^{k}
\right)
\end{align*}
for $k\geq1$.
We then obtain the set of all generators of the cluster algebra of type $A^{(2)}_2$.

\begin{thm}
\label{thm:generators}
The set $\mathcal{X}$ of all generators of the cluster algebra of type $A^{(2)}_2$ are given by
\begin{align*}
&\mathcal{X}
=
\bigcup_{l\in\Z}
\bx_{l}
=
\mathcal{X}_0
\cup
\mathcal{X}_+
\cup
\mathcal{X}_-,
\\
&\mathcal{X}_0
=
\left\{\bx_0\right\}
=
\left\{x_1,x_2\right\},
\\
&\mathcal{X}_+
=
\bigcup_{l>0}
\bx_{l}
=
\left\{
\left.y_1^{k-1}y_2^\frac{2k-3}{4}\rho(k),
y_1^{k}y_2^\frac{2k-1}{4}\left(\rho(k+1)\rho(k)-1\right)\ \right|\ 
k\geq1
\right\},
\\
&\mathcal{X}_-
=
\bigcup_{l<0}
\bx_{l}
=
\left\{
\left.y_1^{k-1}y_2^\frac{2k-3}{4}\widetilde\rho(k),
y_1^{k}y_2^\frac{2k-1}{4}\left(\widetilde\rho(k+1)\widetilde\rho(k)-1\right)\ \right|\ 
k\geq1
\right\},
\end{align*}
where we put
\begin{align*}
\begin{cases}
\DIS
\rho(t)
=
\frac{2y_2^{\frac{1}{4}}}{\lambda^\prime}\left(\frac{y_1x_2+1}{x_1}\sh(\Lambda t)-y_2^{-\frac{1}{2}}x_1\sh(\Lambda (t-1))\right),
\\
\DIS
\widetilde\rho(t)
=
\frac{2y_2^{-\frac{1}{4}}}{\lambda^\prime}\left(y_2^{\frac{1}{2}}x_{1}\sh(\Lambda t)-\frac{y_1x_2+1}{x_{1}}\sh(\Lambda (t-1))\right),
\\
\DIS
\Lambda
=\log\left(\lambda^\prime-\lambda\right)-\log2,
\\
\DIS
\lambda
=
-\frac{\left(y_1x_2+1\right)^2y_2+x_1^4}{y_1y_2^{\frac{1}{2}}x_1^2x_2}.
\end{cases}
\end{align*}
\end{thm}

(Proof)\quad
By substituting $z^0=y_2^{-\frac{1}{4}}x_1$ and $w^0=y_1x_2$ into \eqref{eq:rho}, we obtain $\mathcal{X}_+$.

The set $\mathcal{X}_-$ consists of the cluster variables generated from the initial variables as follows (see figure \ref{fig:binarytree})
\begin{align*}
\bx_0=(x_1,x_2)
&\overset{\mu_2}{\longleftrightarrow}
\bx_{-1}=(x_{1;-1},x_{2;-1})
\overset{\mu_1}{\longleftrightarrow}
\bx_{-2}=(x_{1;-2},x_{2;-2})
\overset{\mu_2}{\longleftrightarrow}
\cdots.
\end{align*}
This sequence is equivalent to
\begin{align}
\bx_1=(x_{1;1},x_{2;1})=(x_{1;1},x_2)
&\overset{\mu_2}{\longleftrightarrow}
\bx_{2}=(x_{1;2},x_{2;2})
\overset{\mu_1}{\longleftrightarrow}
\bx_{3}=(x_{1;3},x_{2;3})
\overset{\mu_2}{\longleftrightarrow}
\cdots
\label{eq:xminus}
\end{align}
with replacing $x_{1;1}$ with $x_1$.
Therefore, we have only to construct the general solution to the dynamical system arising from the sequence \eqref{eq:xminus} of mutations.
The general solution is obtained from the one (see theorem \ref{thm:GS}) to the dynamical system governed by the map $\psi_{\rm vh}$ with replacing $z^0$ with
\begin{align}
\frac{w^0+1}{z^1},
\label{eq:z0replace}
\end{align}
where we use the exchange relation $z^0z^1=w^0+1$.

Actually, in \eqref{eq:rho}, we replace $z^0$ with \eqref{eq:z0replace}, and denote it by $\widetilde\rho(t)$.
We then obtain
\begin{align*}
\widetilde\rho(t)
=
\frac{2}{\lambda^\prime}\left(z^1\sh(\Lambda t)-\frac{w^0+1}{z^1}\sh(\Lambda (t-1))\right).
\end{align*}
Since $\lambda$ is the conserved quantity, it is invariant under the replacement.
By substituting $z^1=y_2^{\frac{1}{4}}x_{1;1}$ and $w^0=y_1x_2$ into $\widetilde\rho(t)$, we get
\begin{align*}
\widetilde\rho(t)
=
\frac{2y_2^{-\frac{1}{4}}}{\lambda^\prime}\left(y_2^{\frac{1}{2}}x_{1;1}\sh(\Lambda t)-\frac{y_1x_2+1}{x_{1;1}}\sh(\Lambda (t-1))\right).
\end{align*}
Finally, replacement of $x_{1;1}$ with $x_1$ completes the proof.
\qed

Similarly, by means of the relation \eqref{eq:zwxyA11}, we have
\begin{align*}
&\bx_{0}
=
\left(
y_2^{\frac{1}{2}}z^0,
y_1^{-\frac{1}{2}}w^{0}
\right),
\quad
\bx_{1}
=
\left(
y_{2}^{-\frac{1}{2}}z^{1},
y_1^{-\frac{1}{2}}w^{0}
\right)
\end{align*}
and
\begin{align*}
&\bx_{2k}
=
\left(
y_1^{k-1}y_2^\frac{2k-3}{2}z^k,
y_1^{\frac{2k-1}{2}}y_2^{k-1}w^{k}
\right),
\quad
\bx_{2k+1}
=
\left(
y_1^{k}y_2^\frac{2k-1}{2}z^{k+1},
y_1^{\frac{2k-1}{2}}y_2^{k-1}w^{k}
\right)
\end{align*}
for $k\geq1$.
We then obtain the set of all generators of the cluster algebra of type $A^{(1)}_1$.

\begin{thm}
\label{thm:generatorsA11}
The set $\mathcal{X}$ of all generators of the cluster algebra of type $A^{(1)}_1$ are given by
\begin{align*}
&\mathcal{X}
=
\bigcup_{l\in\Z}
\bx_{l}
=
\mathcal{X}_0
\cup
\mathcal{X}_+
\cup
\mathcal{X}_-,
\\
&\mathcal{X}_0
=
\left\{\bx_0\right\}
=
\left\{x_1,x_2\right\},
\\
&\mathcal{X}_+
=
\bigcup_{l>0}
\bx_{l}
=
\left\{
\left.y_1^{k-1}y_2^\frac{2k-3}{2}\sigma_1(k),
y_1^{\frac{2k-1}{2}}y_2^{k-1}\sigma_2(k)\ \right|\ 
k\geq1
\right\},
\\
&\mathcal{X}_-
=
\bigcup_{l<0}
\bx_{l}
=
\left\{
\left.\frac{(y_1y_2)^{k-1}}{x_1}\widetilde\sigma_1(k),
\frac{(y_1y_2)^{\frac{2k-1}{2}}}{x_1}\widetilde\sigma_2(k)\ \right|\ 
k\geq1
\right\},
\end{align*}
where we put
\begin{align*}
\begin{cases}
\DIS
\sigma_1(t)
=
y_2^{-\frac{1}{2}}x_1\ch(Mt)-\frac{\nu y_2^{-\frac{1}{2}}x_1+2y_1^{\frac{1}{2}}x_2}{\nu^\prime }\sh(Mt),
\\
\DIS
\sigma_2(t)
=
y_1^{\frac{1}{2}}x_2\ch(Mt)+\frac{\nu y_1^{\frac{1}{2}}x_2+2y_2^{-\frac{1}{2}}x_1}{\nu^\prime }\sh(Mt),
\\
\DIS
\widetilde\sigma_1(t)
=
\left(y_1x_2^2+1\right)\ch(Mt)
-
\frac{\nu y_1x_2^2+\mu+2y_1^{\frac{1}{2}}y_2^{-\frac{1}{2}}x_1x_2}{\nu^\prime }\sh(Mt),
\\
\DIS
\widetilde\sigma_2(t)
=
y_1^\frac{1}{2}x_2\ch(Mt)
+
\frac{\nu y_1^\frac{1}{2}y_2^{-\frac{1}{2}}x_1x_2+2y_1x_2^2+2}{\nu^\prime }\sh(Mt)
\\
\end{cases}
\end{align*}
and
\begin{align*}
\begin{cases}
\DIS
M
=\log\left(\nu^2-2+\nu\nu^\prime \right)-\log2,
\\
\DIS
\nu
=
-\frac{x_1^2+y_1y_2x_2^2+y_2}{y_1^{\frac{1}{2}}y_2^{\frac{1}{2}}x_1x_2}.
\\
\end{cases}
\end{align*}
\qed
\end{thm}

Finally, we give Laurent expressions for several cluster variables in the cluster algebra of type $A^{(2)}_2$ in terms of theorem \ref{thm:generators}.
We then find that they are indeed the Laurent polynomials with positive coefficients in the initial cluster variables $x_1$ and $x_2$  \cite{FZ02,LS15,GHKK18}.
\begin{example}
In order to give cluster variables in the cluster algebra of type $A^{(2)}_2$ by using theorem \ref{thm:generators}, we compute
\begin{align*}
2\sh(\Lambda t)
&=
\left(\frac{-\lambda+\lambda^\prime}{2}\right)^t
-
\left(\frac{-\lambda-\lambda^\prime}{2}\right)^t
\\
&=
\lambda^\prime\left(\frac{-\lambda}{2}\right)^{t-1}\sum_{n=0}^{\lfloor t-1/2\rfloor}\binom{t}{2n+1}\left(\frac{\lambda^\prime}{\lambda}\right)^{2n}.
\end{align*}
We moreover compute
\begin{align*}
\rho(t)
&=
\frac{2}{\lambda^\prime}\left(\frac{w^0+1}{z^0}\sh(\Lambda t)-z^0\sh(\Lambda (t-1))\right)
\\
&=
\left(\frac{-\lambda}{2}\right)^{t-1}
\left\{
\frac{w^0+1}{z^0}\sum_{n=0}^{\lfloor t-1/2\rfloor}\binom{t}{2n+1}\left(\frac{\lambda^\prime}{\lambda}\right)^{2n}
+
\frac{2z^0}{\lambda}\sum_{n=0}^{\lfloor t-2/2\rfloor}\binom{t-1}{2n+1}\left(\frac{\lambda^\prime}{\lambda}\right)^{2n}
\right\}.
\end{align*}
Now, for instance, we give first three $\rho(t)$:
\begin{align*}
\rho(1)
&=
\frac{y_2^{\frac{1}{4}}(y_1x_2+1)}{x_1},
\\
\rho(2)
&=
\frac{(y_1x_2+1)^3y_2+x_1^4}{y_1y_2^{\frac{1}{4}}x_1^3x_2},
\\
\rho(3)
&=
\frac{x_1^4+3y_1^2y_2x_2^2+5y_1y_2x_2+2y_2}{y_1^2y_2^{\frac{3}{4}}x_1x_2^2}
+
\frac{(y_1x_2+1)^5y_2^\frac{5}{4}}{y_1^2x_1^5x_2^2}.
\end{align*}

Then, by means of theorem \ref{thm:generators}, the following 10 cluster variables are explicitly computed
\begin{align*}
&x_{1;1}
=
x_{1;2}
=
y_2^{-\frac{1}{4}}\rho(1)
=
\frac{y_1x_2+1}{x_1},
\\
&x_{1;3}
=
x_{1;4}
=
y_1y_2^{\frac{1}{4}}\rho(2)
=
\frac{x_1^2}{x_2^2}
+
\frac{(y_1x_2+1)^3y_2}{x_1^2x_2},
\\
&x_{1;5}
=
x_{1;6}
=
y_1^2y_2^{\frac{3}{4}}\rho(3)
=
\frac{x_1^3}{x_2^2}
+
\frac{3y_1^2y_2}{x_1}
+
\frac{5y_1y_2}{x_1x_2}
+
\frac{2y_2}{x_1x_2^2}
+
\frac{(y_1x_2+1)^5y_2^2}{x_1^5x_2^2}
\end{align*}
and
\begin{widetext}
\begin{align*}
x_{2:2}
&=
x_{2;3}
=
y_1(\rho(2)\rho(1)-1)
=
\frac{1}{x_2}
+
\frac{y_2(y_1x_2+1)^4}{x_1^4x_2},
\\
x_{2:4}
&=
x_{2;5}
=
y_1^3y_2(\rho(3)\rho(2)-1)
\\
&=
\frac{y_2+x_1^{4}}{x_2^3}
+
\frac{y_1x_2+1}{x_2^2}\left\{3y_1y_2+\left(1+\frac{(y_1x_2+1)^3y_2}{x_1^4}\right)\left(3y_1^2+\frac{2y_1}{x_2}+\frac{(y_1x_2+1)^4y_2^2}{x_1^4x_2}\right)\right\}
.
\end{align*}
\end{widetext}
Indeed, these are the Laurent polynomials with positive coefficients in the initial cluster variables $x_1$ and $x_2$.
\end{example}

\section{Concluding remarks}
\label{sec:CONCL}
From the seed mutations of type $A^{(2)}_2$, we constructed the discrete integrable system on the singular quartic curve $\gamma_\lambda$ governed by the quartic birational map $\psi_{\rm vh}$.
Singularity of the curve $\gamma_\lambda$ was then resolved by blowing-up the curve at the singular point $\mathfrak{p}$, and the non-singular conic $\delta_\lambda$ was obtained.
The map $\psi_{\rm vh}$ on the singular curve $\gamma_\lambda$ was simultaneously transformed into the one $\phi_{\rm dv}$ on the conic $\delta_\lambda$ by the blowing-up.
Remark that the blowing-up plays a central role in of our method because it enables us to linearize the birational map $\psi_{\rm vh}$.
The conic $\delta_\lambda$ thus obtained is nothing but the invariant curve of the QRT map $\phi_{\rm vh}$ arising from the seed mutations of type $A^{(1)}_1$.
Moreover, these two birational maps $\phi_{\rm dv}$ and $\phi_{\rm vh}$ commute with each other on the conic $\delta_\lambda$ and they are linearized simultaneously by using the conic. 
Finally, we respectively solved the initial value problems of the integrable systems governed by the birational maps $\psi_{\rm vh}$ and $\phi_{\rm dv}$, and we respectively presented the sets of all cluster variables generating the cluster algebras of types $A^{(1)}_1$ and $A^{(2)}_2$ in terms of the general solutions to the integrable systems.  

The discrete integrable system governed by the birational map $\phi_{\rm dv}$ is equivalent to the system subject to the following difference equation for some $\beta$:
\begin{align}
x_{n-1}x_nx_{n+1}=x_{n-1}+(x_n)^{\beta-1}+x_{n+1}
\quad
(\beta\in\mathbb{N}).
\label{eq:2dds}
\end{align}
Actually, by eliminating $v^t$ form \eqref{eq:tildepsi}, we obtain the difference equation \eqref{eq:2dds} for $\beta=4$.
The difference equation \eqref{eq:2dds} is arising from seed mutations of a cluster algebra with a special choice of the initial seed.
It seems to be very interesting to investigate the difference equation \eqref{eq:2dds} since it is integrable for $\beta\leq4$, while it exhibits chaos for $\beta\geq5$.
Actually, we can show that the algebraic entropy of \eqref{eq:2dds} with $\beta>5$ is a positive number.
Moreover, it fails the singularity confinement test \cite{GRP91,HV98} for any $\beta\geq4$.
When $\beta=4$ this fact is consistent with linearizability of the map $\phi_{\rm dv}$, while it reinforces non-integrability of the difference equation \eqref{eq:2dds} when $\beta>5$ .
We will report integrable and non-integrable properties of \eqref{eq:2dds} precisely in a forthcoming paper.

\begin{acknowledgments}
This work was partially supported by JSPS KAKENHI Grant Number 26400107.
\end{acknowledgments}

\appendix

\section{How do we find invariant curves?}
\label{sec:App}
Let us tropicalize  \cite{RST05} (or ultradiscretize \cite{TTMS96}) the birational map $\psi_{\rm vh}$ arising from the seed mutations of type $A^{(2)}_2$.
We denote the piecewise linear map thus obtained by $\Psi_{\rm vh}$:
\begin{align}
(Z^t,W^t)
&\mapsto
(Z^{t+1},W^{t+1})
=
\left(
\min\left[W^t,0\right]-Z^t,
\min\left[4Z^{t+1},0\right]-W^t
\right).
\label{eq:bmA22UD}
\end{align}
The tropical variables are denoted by the capital letters.
The variable transformation
\begin{align*}
z^t=\exp\left(-\frac{Z^t}{\varepsilon}\right),
\quad
w^t=\exp\left(-\frac{W^t}{\varepsilon}\right)
\end{align*}
for $\varepsilon>0$ and the limiting procedure $\varepsilon\to0$ reduces \eqref{eq:bmA22} to \eqref{eq:bmA22UD} \cite{TTMS96}.
Note that the operations $\times$ and $+$ reduce to $+$ and $\min$, respectively.
Note also that $1$, the unit of $\times$, reduces to $0$, the unit of $+$.

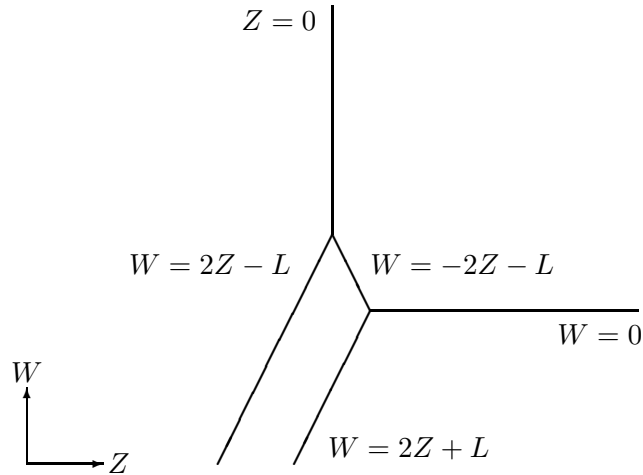
\begin{figure}[htb]
\centering
\unitlength=.04in
\def\arraystretch{1.0}
\begin{picture}(70,70)(-33,-30)
\put(-40,-30){\vector(0,1){10}}
\put(-40,-30){\vector(1,0){10}}
\thicklines
\put(0,0){\line(0,1){30}}
\put(0,0){\line(-1,-2){15}}
\put(0,0){\line(1,-2){5}}
\put(5,-10){\line(-1,-2){10}}
\put(5,-10){\line(1,0){35}}
\put(-28,-30){\makebox(0,0){$Z$}}
\put(-40,-18){\makebox(0,0){$W$}}
\put(-7,28){\makebox(0,0){$Z=0$}}
\put(35,-13){\makebox(0,0){$W=0$}}
\put(17,-4){\makebox(0,0){$W=-2Z-L$}}
\put(10,-28){\makebox(0,0){$W=2Z+L$}}
\put(-16,-4){\makebox(0,0){$W=2Z-L$}}
\end{picture}
\caption{
The tropical curve $\Gamma_L$ defined by the tropical polynomial $F(Z,W)$.
The curve $\Gamma_L$ is the invariant curve of the dynamical system governed by the piecewise linear map $\Psi_{\rm vh}$.
}
\label{fig:A22UDIV}
\end{figure}

Numerical experiments suggest us that the invariant curve of the piecewise linear map $\Psi_{\rm vh}$ is the tropical curve $\Gamma_L$ defined by the tropical polynomial \cite{RST05}
\begin{align*}
F(Z,W):=\min\left[2W,0,2Z+W+L,4Z\right],
\end{align*}
where $L\in\R$ is a parameter (see figure \ref{fig:A22UDIV}).

\begin{proposition}
The tropical curve $\Gamma_L$ is the invariant curve of the dynamical system governed by the piecewise linear map $\Psi_{\rm vh}$ and $L=\min\left[2W^t,0,4Z^t\right]-2Z^t-W^t$ is the conserved quantity of the system.
\end{proposition}

(Proof)\quad
First we denote the horizontal flip and the vertical flip by
\begin{align*}
&\Psi_{\rm h}:(Z^t,W^t)\mapsto(Z^{t+1},W^t)
=
\left(
\min\left[W^t,0\right]-Z^t,
W^t
\right),
\\
&\Psi_{\rm v}:(Z^t,W^t)\mapsto(Z^{t},W^{t+1})
=
\left(
Z^t,
\min\left[4Z^{t},0\right]-W^t
\right),
\end{align*}
respectively.
Then we have the decomposition $\Psi_{\rm vh}=\Psi_{\rm v}\circ\Psi_{\rm h}$ of the piecewise linear map.

Let $\ell_1,\ell_2,\ldots,\ell_5$ be the line segments or the half lines respectively defined as follows
\begin{align*}
&\ell_1
=
\left\{
(Z,W)\in\R^2\ |\ 
W=2Z-L,\ W\leq0
\right\},
\\
&\ell_2
=
\left\{
(Z,W)\in\R^2\ |\ 
W=2Z-L,\ Z\leq0,\ W\geq0
\right\},
\\
&\ell_3
=
\left\{
(Z,W)\in\R^2\ |\ 
W=-2Z-L,\ Z\geq0,\ W\geq0
\right\},
\\
&\ell_4
=
\left\{
(Z,W)\in\R^2\ |\ 
W=2Z+L,\ Z\geq0,\ W\leq0
\right\},
\\
&\ell_5
=
\left\{
(Z,W)\in\R^2\ |\ 
W=2Z+L,\ Z\leq0
\right\}.
\end{align*}
Note that $\ell_1\cup\ell_2$, $\ell_3$ and $\ell_4\cup\ell_5$ are the edges of the curve $\Gamma_L$ defined by $W=2Z-L$, $W=-2Z-L$ and $W=2Z+L$, respectively (see figure \ref{fig:A22UDIV}).

Let $(Z^0,W^0)$ be a point on $\R^2$.
Then there is a unique curve $\Gamma_L$ passing through $(Z^0,W^0)$.
More precisely, we have three cases; the point $(Z^0,W^0)$ is on
\begin{align*}
\begin{cases}
\ell_1\cup\ell_2&\mbox{if $2Z^0\leq\min\left[W^0,0\right]$,}\\
\ell_3&\mbox{if $0\leq\min\left[2Z-0,W^0\right]$,}\\
\ell_4\cup\ell_5&\mbox{if $W^0\leq\min\left[2Z^0,0\right]$.}\\
\end{cases}
\end{align*}
Thus the value of $L$ is uniquely fixed:
\begin{align*}
L
=
\min\left[2W^0,0,4Z^0\right]-2Z^0-W^0.
\end{align*}
We easily see that the value of $L$ is non-positive.

Let $t>0$ be a positive integer. 
Assume that the point $(Z^s,W^s)$ is on the curve $\Gamma_L$ for $0\leq s\leq t$.
Then we see that the horizontal flip $\Psi_{\rm h}$ maps a point on $\ell_1$ into a point on $\ell_4\cup\ell_5$.
Actually, assume $(Z^t,W^t)\in\ell_1$ we have
\begin{align*}
2Z^{t+1}+L
&=
2\min\left[W^t,0\right]-2Z^t+L
\\
&=
2W^t-2Z^t+L
\\
&=
W^t,
\end{align*}
where we use $L=2Z^t-W^t$ since $(Z^t,W^t)\in\ell_1$.
Thus, we have $(Z^{t+1},W^t)\in\ell_4\cup\ell_5$.
We similarly obtain the following correspondence by applying the horizontal and the vertical flips, respectively:
\begin{align*}
&\Psi_{\rm h}:\ell_1\longleftrightarrow\ell_4\cup\ell_5,
\quad
\Psi_{\rm h}:\ell_2\longleftrightarrow\ell_3,
\\
&\Psi_{\rm v}:\ell_1\cup\ell_2\longleftrightarrow\ell_5,
\quad
\Psi_{\rm v}:\ell_3\longleftrightarrow\ell_4.
\end{align*}
Induction on $t$ leads to the conclusion.
It is clear that $L=\min\left[2W^t,0,4Z^t\right]-2Z^t-W^t$ is the conserved quantity of the map $\Psi_{\rm vh}$.
\qed

Now let us consider inverse procedure of tropicalization.
First we reduce the tropical polynomial $F(Z,W)$ to the equivalent one
\begin{align*}
F(Z,W)=\min\left[2\min\left[W,0\right],2Z+W+L,4Z\right].
\end{align*}
Replace the operations $\min$ and $+$ in $F(Z,W)$ with $+$ and $\times$, respectively.
Then the tropical polynomial $F(Z,W)$ changes into
\begin{align*}
f(z,w)
=
(w+1)^2+\lambda z^2w+z^4,
\end{align*}
where the tropical variables denoted by capital letters are replaced with the ones denoted by small letters and the parameter $L$ with $\lambda$.
This is nothing but the defining polynomial of the invariant curve $\gamma_\lambda$ of the birational map $\psi_{\rm vh}$ arising from the seed mutations of type $A^{(2)}_2$ (see \eqref{eq:ICA22}).

\begin{remark}
The piecewise linear map $\Psi_{\rm vh}$ is linearized on the tropical Jacobian $J(\Gamma_L)$ of the tropical curve $\Gamma_L$ \cite{Nobe08}.
Namely, $\Psi_{\rm vh}$ is mapped into the translation with length $3L$ on $J(\Gamma_L)$ by the tropical Abel-Jacobi map.
\end{remark}

\nocite{*}
\providecommand{\noopsort}[1]{}\providecommand{\singleletter}[1]{#1}%

\end{document}